\documentclass[twocolumn,showpacs,preprintnumbers]{revtex4}
\usepackage{graphicx}
\usepackage{dcolumn}
\usepackage{bm}

\begin{document}

\title{Interference of quantum channels in single photon interferometer}
\author{Guo-Yong Xiang\thanks{%
Electronic address: gyxiang@mail.ustc.edu.cn}, Jian Li, and Guang-Can Guo%
\thanks{%
Electronic address: gcguo@ustc.edu.cn}}
\address{Key Laboratory of Quantum Information, University of Science and Technology\\
of China, CAS, Hefei 230026, People's Republic of
China\bigskip\bigskip }


\begin{abstract}
We experimently demonstrate the interference of dephasing quantum channel
using single photon Mach-Zender interferometer. We extract the information
inaccessible to the technology of quantum tomography. Further, We introduce
the application of our results in quantum key distribution.
\end{abstract}

 \maketitle

Macroscopic quantum systems can never be isolated from their environments.
It leads to decoherence which destroys superpositions. And when a qubit
transmits through a quantum channel, the interaction between qubit and
quantum channel is inevitable. The decoherence in quantum channel affects
the distance and quantity of quantum information transmitting. So it is
important to know what happen when quantum information transmit through
noisy quantum channel. The technology of quantum process tomography\cite%
{Childs01,Jezek02} can be used to character the quantum channels.

But J. Aberg\cite{aberg03} find that we can not specify the action of the
sinultaneous operation of both maps although we known the individual quantum
channels. It is said that when a superposition state pass through two
quantum channels, we can not know the information of output state exactly by
using the technology of quantum process tomography. Single particle
interference can help us extract information inaccessible to conventional
process tomography. D. K. L. Oi have given a measure of coherent fidelity,
the maximum interference visibility, and the closest unitary operator to a
given physical process under this measure\cite{dan03}.

Here, We give an interference visibility of two quantum processes which have
same environment degree and carry out an experiment to demonstrate it. The
environment qubit is the time qubit from birefrigence of quartz crystal in
the experiment, i.e. quantum channels we used is the dephasing channel. We
find that there are plentiful information of inteference which is the
information inaccessible to conventional process tomography\cite%
{Childs01,Jezek02}.

When a single qubit state transmits through two quantum channels (Fig. 1),
how can we known the output state? The technology of quantum tomography can
obtain the densities of output states in each paths. But the whole density
of the output state can not be fixed, i.e. there is other information which
have not been extracted. D. K. L. Oi\cite{dan03} shows that single particle
Mach-Zender inereference can help us. Different visibilities show quantum
information not presented in the two individual quantum channel. When the
different environment degree ($E$ and $F$) appended to the operations of the
upper and lower arms, D. K. L. Oi presents the interference patterns as
\begin{equation}
Tr[u_{0}^{+}v_{0}\rho ],
\end{equation}%
where $\rho $ is the input state, $u_{0}$ and $v_{0}$ are the first Kraus
operators for the quantum processes $U$ and $V$ in upper and lower arms. If
the input state is the maximally mixed state, the interference pattern
depends on $\frac{1}{d}Tr[u_{0}^{+}v_{0}]$.

In Eq. 1, The interference patterns only depend on the first Kraus operators
$u_{0}$ and $v_{0}$. But it find that when the environment degree is same to
the operations of the both arms (Fig. 2), the interference pattern will
depend on the four Kraus operators $u_{i}$ and $v_{i}$. The beamsplitters in
Fig. 2 and the phase shifter are modeled by the unitary operators $U_{b}$
and $U_{p}$ respectively,%
\begin{equation}
U_{b}=\frac{1}{\sqrt{2}}\left(
\begin{array}{cc}
1 & 1 \\
-1 & 1%
\end{array}%
\right) ,U_{p}=\left(
\begin{array}{cc}
1 & 0 \\
0 & e^{i\phi }%
\end{array}%
\right) .
\end{equation}%
The original state $\rho _{in}=\left\vert 0\right\rangle \left\langle
0\right\vert \otimes \rho $ of the system on internal Hilbert space and the
two-dimensional Hilbert space of path degree is evolved as

\begin{equation}
\rho _{in}\mapsto U_{b}(\left\vert 0\right\rangle \left\langle
0\right\vert U+\left\vert 1\right\rangle \left\langle 1\right\vert
V)U_{p}U_{b}\rho _{in}U_{b}^{\dagger }U_{p}^{\dagger }
(\left\vert 0\right\rangle \left\langle 0\right\vert U^{\dagger
}+\left\vert 1\right\rangle \left\langle 1\right\vert V^{\dagger
})U_{b}^{\dagger }
\end{equation}%
where $\left\vert 1\right\rangle $ and $\left\vert 0\right\rangle $
represent the upper and lower path. The probability of finding the particle
in the horizontal direction, i.e. in the $\left\vert 0\right\rangle $ state,
is
\begin{equation}
P_{\left\vert 0\right\rangle }(\phi )=\frac{1}{2}(1+Re{e^{i\phi
}Tr[U^{+}V\rho \otimes \left\vert e_{0}\right\rangle \left\langle
e_{0}\right\vert ]}).
\end{equation}

So $P_{\left\vert 0\right\rangle }(\phi )$ is decided by
\begin{figure}[tbh]
\includegraphics[width=8cm]{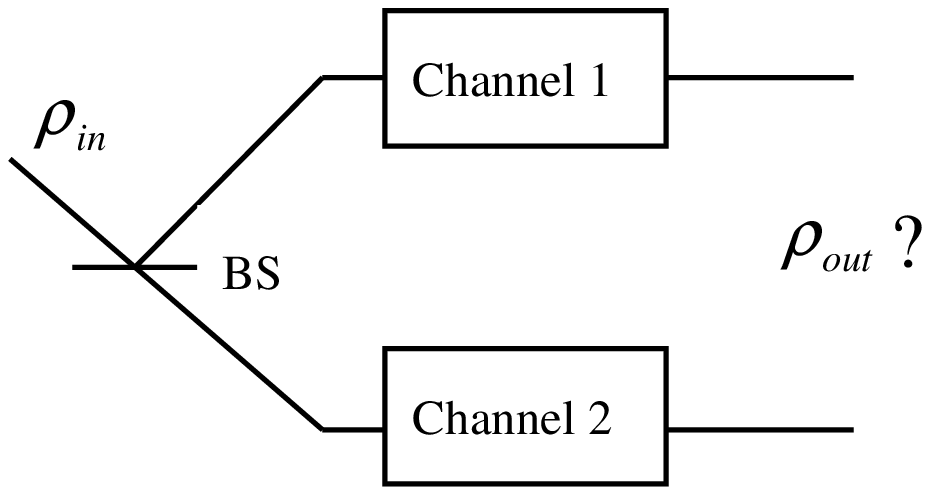}
\caption{After the beam splitter(BS), the input state $\protect\rho $ will
transmit through a coherent superposition of quantum channel 1 and quantum
channel 2. The output state $\protect\rho _{out}$ can not be determined by
conventional process tomography}
\end{figure}
\begin{equation}
Tr[U^{+}V\rho \otimes \left\vert e_{0}\right\rangle \left\langle
e_{0}\right\vert ]=\sum\limits_{i}Tr[u_{i}^{+}v_{i}\rho ],
\end{equation}%
\begin{figure}[tbh]
\includegraphics[width=8cm]{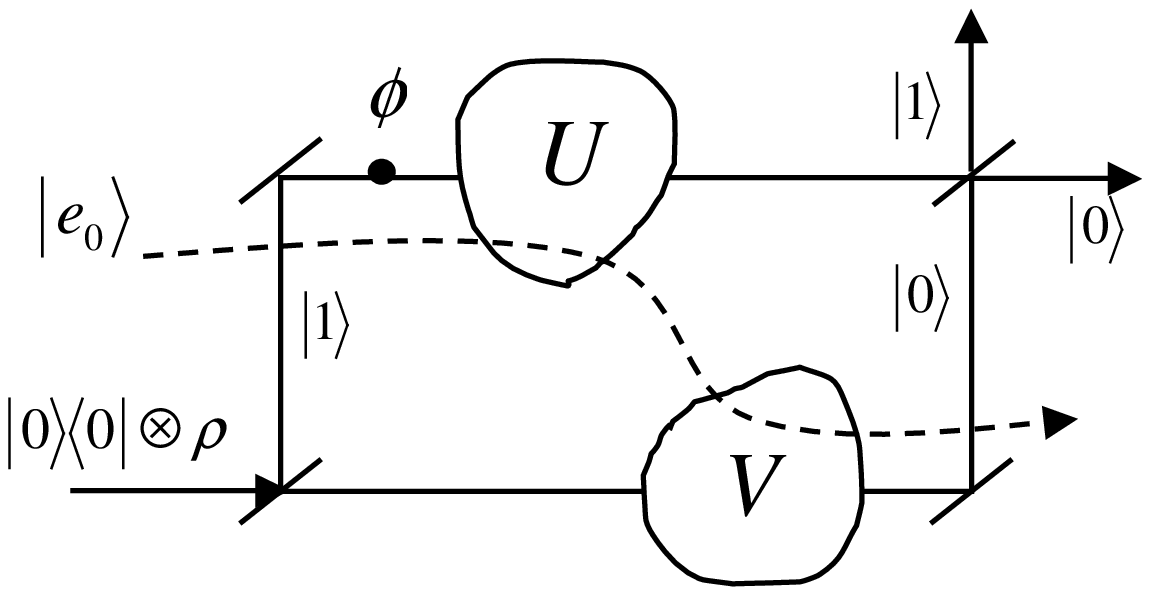}
\caption{Interference of two quantum channels. The environment degree of the
two channels is same. After the second BS, the environment qubit will be
traced out}
\end{figure}
where
\begin{equation}
\{u_{i}\}=\{\left\langle e_{i}\right\vert U\left\vert e_{0}\right\rangle
\},\{v_{j}\}=\{\left\langle e_{j}\right\vert V\left\vert e_{0}\right\rangle
\}
\end{equation}%
are the Kraus operators of $U$ and $V$. Where $\left\{ \left\vert
e_{i}\right\rangle \right\} $ and $\left\{ \left\vert e_{j}\right\rangle
\right\} $ are the orthornomal bases of E, $\left\vert e_{0}\right\rangle $
is the initial state of E. Specially, If the input state is the maximally
mixed state, the interference pattern depends on $\frac{1}{d}%
Tr[u_{i}^{+}v_{i}]$, which is the interference patterns of quantum channels.

So the interference patterns determined by all four kraus operators of $U$
and $V$ and their relative phase. It will give richer interference pattern
than that Eq. 1 gives us.

The visibility of the intereference pattern is the effects of the
indistinguishability of the two paths that the particle transmitted through.
According to the intereference pattern and visibility, it can be determined
whether the two quantum processes are identical or different (see \cite%
{ander03} for related problem). Because of the birefringence of ordinary
light ($o$ light) and extra-ordinary light ($e$ light) in BBO crystal, we
choose the time degrees of freedom of photon passing through a BBO crystal
to be the environment qubit ($\left\vert t_{o}\right\rangle $ to for
ordinary light and $\left\vert t_{e}\right\rangle $ for extraordinary
light). There are two birefrigence crystals in the upper and lower arms of
Mach-Zender interferometer respectively (see Fig. 3). The Kraus operators of
two arms can be represented by%
\begin{eqnarray}
\{u_{m}\} &=&\{\left\langle a_{i}|b_{j}\right\rangle \left\vert
a_{i}\right\rangle \left\langle b_{j}\right\vert \},  \nonumber \\
\{v_{n}\} &=&\{\left\langle a_{i}^{\prime }|b_{j}^{\prime }\right\rangle
\left\vert a_{i}^{\prime }\right\rangle \left\langle b_{j}^{\prime
}\right\vert \}
\end{eqnarray}%
Where $m,n=0,1,2,3$, and $a_{i},b_{j},a_{i}^{\prime },b_{j}^{\prime }$ are
the angles of the fast axis of the crystals relative to horizontal
direction; $i,j=o,e$, and $\{\left\vert a_{o}\right\rangle =\cos a\left\vert
H\right\rangle +\sin a\left\vert V\right\rangle ,\left\vert
a_{e}\right\rangle =-\sin a\left\vert H\right\rangle +\cos a\left\vert
V\right\rangle \},\{\left\vert b_{o}\right\rangle =\cos b\left\vert
H\right\rangle +\sin b\left\vert V\right\rangle ,\left\vert
b_{e}\right\rangle =-\sin b\left\vert H\right\rangle +\cos b\left\vert
V\right\rangle \}$ are orthogonal basis respectively. Here, $m,n$ are
defined by the sequence of the four pulse following the second quartz
crystal in the two arms respectively. For example, $u_{0}=\left\langle
a_{0}|b_{0}\right\rangle \left\vert a_{0}\right\rangle \left\langle
b_{0}\right\vert ,u_{1}=\left\langle a_{0}|b_{e}\right\rangle \left\vert
a_{0}\right\rangle \left\langle b_{e}\right\vert ,u_{2}=\left\langle
a_{e}|b_{0}\right\rangle \left\vert a_{e}\right\rangle \left\langle
b_{0}\right\vert ,u_{3}=\left\langle a_{e}|b_{e}\right\rangle \left\vert
a_{e}\right\rangle \left\langle b_{e}\right\vert $.

According to Eq. 4, the interference patterns are determined by

\begin{equation}
ve^{i\phi }=\sum\limits_{i}Tr[u_{m}^{+}v_{n}\rho ].
\end{equation}%
\begin{figure}[tbh]
\includegraphics[width=8cm]{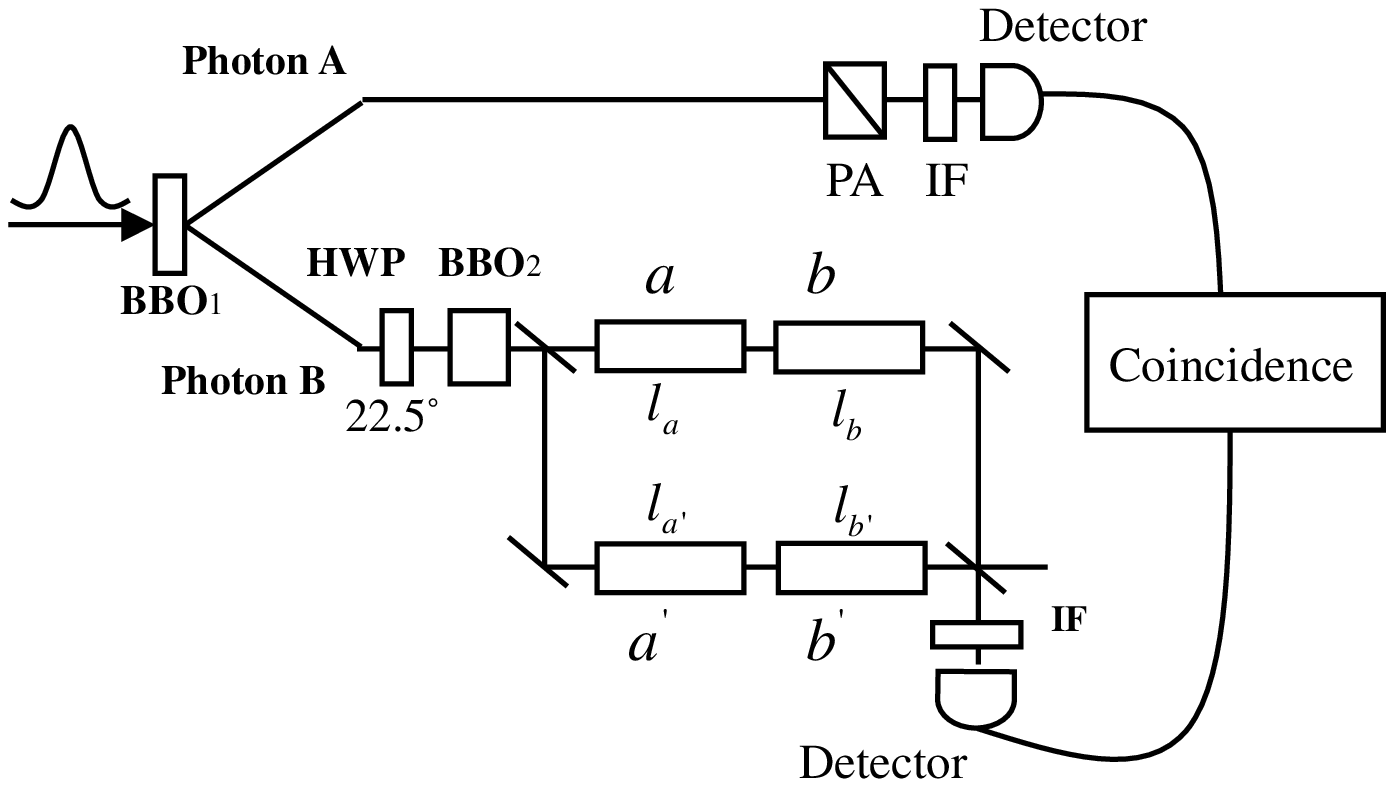}
\caption{Experimental setup for the interference of quantum channels. PA
represents polarization analyzer and IF represents interference filter; HWP
are half waveplates.}
\end{figure}
The experimental setup is represented in Fig. 3. A pulse of ultraviolet (UV)
light pass through a BBO crystal ($1.0mm$, cut for type-I phase match) . The
UV pulse is frequency-doubled pulse (less than $200fs$ with $82MHz$
repetition and $390nm$ center-wavelength) from a mode-locked Ti: sapphire
laser (Tsunami by Spectra-Physics). Through the SPDC process, photon pairs
are generated with $780nm$ center-wavelength. By detecting one photon of the
pairs (with single photon detector after a $4nm$ FWHM interference filter at
$780nm$), the other one (photon $1$) can be prepared into any polarization
state\cite{kwiat03,xiang05} to be sent into Mach-Zender interferometer.

After a half-wave plate fixed $22.5^{0}$ and a $5.0mm$ thick BBO crystal
(After which, the separation of wavepackets between H(o)- and V(e)-polarized
light is about $580\mu m$) and Because the coherent length of the wavepacket
is about $150\mu m$ ($4nm$ FWHM interference filter is inserted before each
detector), Photon $1$ is prepared in the maximally mixed state. Then it sent
into Mach-Zender interferometer (Fig. 3). There are two quarz crystals in
the upper and lower arms respectively. The two short ones ($l_{1}$) separate
H(o)- and V(e)-polarized light $190\lambda $ (about $150\mu m$), and the two
longer ones ($l_{2}$) separate H(o)- and V(e)-polarized light $398\lambda $
(about $310\mu m$). The angles of their
\begin{figure}[tbh]
\includegraphics[width=8cm]{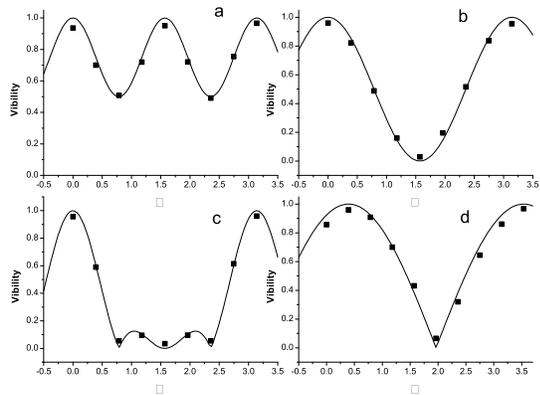}
\caption{The experimental results of interference of quantum
channel.}
\end{figure}
optical axes relative to horizontal plane are $a,b,a^{\prime },b^{\prime }$
(see Fig. 3). Then beams from the two arms interact at the second beam
splitter, and photon $1$ is detected by single photon detector below
interferometer after a $4nm$ FWHM interference filter. The signals from two
detectors are coincided within a $5ns$ timing window by using a coincidence
counter (EG\&G, TAC/SCA).

The maps according to quantum channel are changed by adjusting the angles ($%
a,b,a^{\prime },$ and $b^{\prime }$) and the arrangement of four quartz
crystals. we will observe the visibility of interference of different
quantum channels. 1), we choose $l_{a}=l_{b^{\prime }}=l_{1}$, $%
l_{b}=l_{a^{\prime }}=l_{2}$, $b=a^{\prime }=0$, and $a=b^{\prime }=\beta $,
then the visibility is $v=1-\frac{\sin ^{2}(2\beta )}{2}$ which is always
more than $50\%$ (Fig. 4a); 2), $l_{a}=l_{b^{\prime }}=l_{1}$, $%
l_{b}=l_{a^{\prime }}=l_{2}$, $a=a^{\prime }=0$, and $b=b^{\prime }=\beta $,
then the visibility is $v=\cos ^{2}\beta $ (Fig. 4b); 3), $%
l_{b}=l_{b^{\prime }}=l_{1}$, $l_{a}=l_{a^{\prime }}=l_{2}$, $b=a^{\prime
}=0 $, and $a=b^{\prime }=\beta $, then the visibility is $v=\cos ^{2}\beta
\cos (2\beta )$ (Fig. 4c); 4), There is a half-wave plate in the upper and
lower arms, \textit{i.e. }$l_{a}=l_{b^{\prime }}=l_{b}=l_{a^{\prime }}=0$,
and the angles of the one in upper and lower arm are fixed in $\frac{\pi }{8}
$and $\beta $ respectively, then the visibility is $v=\left\vert \cos (\beta
-\frac{\pi }{8})\right\vert $ (Fig. 4d). Because, to the maximally mixed
states input, the outpt state are still maximally states after the quartz
crystals in both arms and the fidelity of the output states of both arms are
always $100\%$ which are independent of the maps in the arms\cite{xiang05},
the change of the visibilities according to $\beta $ is the information
inaccessible to conventional process tomography.

\begin{figure}[tbh]
\includegraphics[width=8cm]{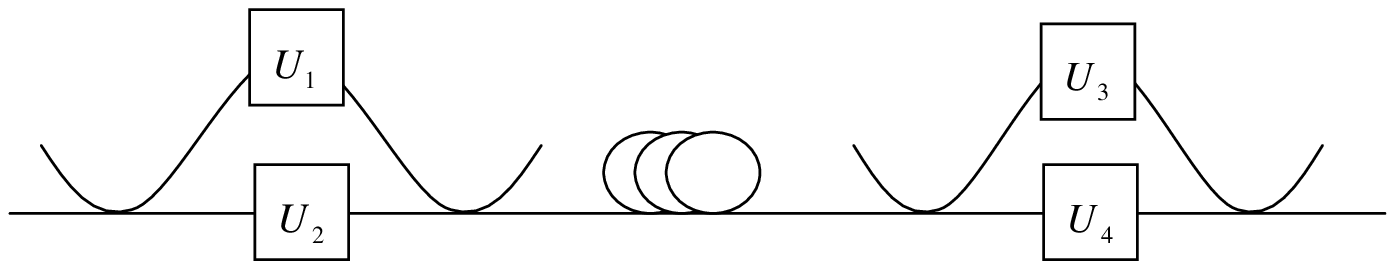}
\caption{The schematic representation of QKD scheme based on unbalanced
fiber Mach--Zender interferometers.}
\end{figure}

Our results can be used to explain the low visibility of Mach--Zender
interferometer in QKD system(\cite{mo05,mul97,boi04,hon04}). A typical fiber
QKD scheme is based on unbalanced fiber Mach--Zender interferometers (Fig.
5). Here we suppose the common quantum channel between two unbalanced
Mach--Zender interferometers was identity. So this QKD system can be
simplified to one Mach--Zender interferometer (see Fig. 3),and $U_{i}$ ($%
i=1,2,3,4$) correspond to four quartz crystals in our experiment. Eq. [5]
gives the visibility of any input sate. Our further work will demonstrate
the visibility of QKD scheme when the common quantum channel between two
unbalanced Mach--Zender interferometers is not identity

In summary, we have demonstrated the interference of quantum channels single
photon Mach-Zender interferometer. Our results present the information
inaccessible to the techonology of quantum process tomography. This work can
lead to further investigation into the phase between operations and
structure and geometry of the CP maps.\bigskip

The authors thank Zheng-Fu Han, Zheng-Wei Zhou and Yun-Feng Huang for
interesting and helpful discussion. This research was funded by National
Fundamental research Program (2001CB309300), the Innovation funds from
Chinese Academy of Sciences KGCX2-SW-112 , and National Natural Science
Foundation of China.

\end{document}